\documentclass{emulateapj}

\newcommand{\f}{\frac}
\newcommand{\p}{\partial}

\newcommand{\bvf}{Brunt--V\"ais\"al\"a frequency}
\newcommand{\sgn}{\mbox{$\mathrm{sgn}$}}
\newcommand{\tr}{OGLE-TR-56b}

\begin{document}

\title{Tidal dissipation in rotating solar-type stars}

\shorttitle{Tidal dissipation in rotating solar-type stars}
\author{G. I. Ogilvie\altaffilmark{1,2} and D. N. C. Lin\altaffilmark{2}}
\altaffiltext{1}{Department of Applied Mathematics and Theoretical
Physics, University of Cambridge, Centre for Mathematical Sciences,
Wilberforce Road, Cambridge CB3 0WA, UK; {\tt gio10@cam.ac.uk}}
\altaffiltext{2}{UCO/Lick Observatory, University of California, Santa Cruz,
CA 95064; {\tt lin@ucolick.org}}
\shortauthors{Ogilvie \& Lin}

\begin{abstract}
  We calculate the excitation and dissipation of low-frequency tidal
  oscillations in uniformly rotating solar-type stars.  For tidal
  frequencies smaller than twice the spin frequency, inertial waves
  are excited in the convective envelope and are dissipated by
  turbulent viscosity.  Enhanced dissipation occurs over the entire
  frequency range rather than in a series of very narrow resonant
  peaks, and is relatively insensitive to the effective viscosity.
  Hough waves are excited at the base of the convective zone and
  propagate into the radiative interior.  We calculate the associated
  dissipation rate under the assumption that they do not reflect
  coherently from the center of the star.  Tidal dissipation in a
  model based on the present Sun is significantly enhanced through the
  inclusion of the Coriolis force but may still fall short of that
  required to explain the circularization of close binary stars.
  However, the dependence of the results on the spin frequency, tidal
  frequency, and stellar model indicate that a more detailed
  evolutionary study including inertial and Hough waves is required.
  We also discuss the case of higher tidal frequencies appropriate to
  stars with very close planetary companions.  The survival of even
  the closest hot Jupiters can be plausibly explained provided that
  the Hough waves they generate are not damped at the center of the
  star.  We argue that this is the case because the tide excited by a
  hot Jupiter in the present Sun would marginally fail to achieve
  nonlinearity.  As conditions at the center of the star evolve,
  nonlinearity may set in at a critical age, resulting in a relatively
  rapid inspiral of the hot Jupiter.
\end{abstract}

\keywords{binaries: close --- planetary systems --- stars:
oscillations ---hydrodynamics --- waves}

\section{Introduction}

\subsection{Tidal interactions involving solar-type stars}

The tidal interaction between a star and an orbiting companion can
lead to a significant evolution of the system over astronomical
timescales if the orbit is sufficiently close.  When the star
experiences a periodically varying gravitational potential, an
oscillatory tidal disturbance is generated in the fluid.  Dissipation
of the tide, by whatever mechanism, is directly associated with a
secular transfer of angular momentum between the spin and the orbit as
well as a loss of energy from the system.  All of the spin and orbital
parameters therefore evolve as a result of tidal dissipation.

In binary stars the tendency of tidal evolution is usually towards a
double synchronous state in which the orbit is circular and both stars
spin at the same rate as the orbit \citep[e.g.][]{H81}.  In this state
each star experiences only a steady tidal distortion and no further
dissipation or tidal evolution occurs.  Observations of stellar
populations of different ages provide clear evidence for the ongoing
tidal circularization of close binaries, because in the older samples
circular orbits are found for a wider range of orbital periods
\citep{MM05}.  Measurements of spin are more limited but do provide
evidence of tidal synchronization \citep{MMS06}.

Many of the known extrasolar planets orbit sufficiently close to their
host stars to allow significant tidal evolution.  The closest planets,
commonly known as ``hot Jupiters'' or ``hot Neptunes'', have circular
orbits, probably mainly as a result of the tidal dissipation within
the planet rather than that within the star.  Although these planets
are expected to be tidally synchronized, the host stars, which are
typically similar to the Sun in mass and age, are not.  The reason is
that the orbital moment of inertia of a hot Jupiter or Neptune is at
most comparable to the spin moment of inertia of a solar-type star.
The tidal torque therefore leads to orbital migration accompanied by a
modest change in the spin rate of the star, which is in any case under
the control of magnetic braking.

The direction of orbital migration is away from the corotation radius
of the star.  When the system is young and the star rapidly rotating,
the inward migration of a hot Jupiter driven by its interaction with
the protoplanetary disk may therefore be halted by the tide it raises
in the star \citep{LBR96}.  Once the star has spun down, however,
tidally driven inspiral threatens the survival of hot Jupiters such as
\tr\ \citep{S03}.

\subsection{Timescales of tidal evolution}
\label{s:timescales}

The efficiency of tidal dissipation is often parametrized by a
dimensionless quality factor $Q$ \citep[e.g.][]{GS66}, which reflects
the fact that the star undergoes a forced oscillation and dissipates a
small fraction of the associated energy during each oscillation
period.  Since $Q$ always appears in the theory in the combination
\begin{equation}
  Q'=\f{3Q}{2k_2},
\end{equation}
where $k_2$ is the potential Love number of order~2 (a dimensionless
measure of the interior density profile of the star), it is more
convenient to use this combination.  (For a homogeneous body without
rigidity, $k_2=3/2$ and $Q'=Q$.)

The tidal potential experienced by the star can be written as a sum of
rigidly rotating components in the form of solid spherical harmonics,
\begin{equation}
  \mathrm{Re}\left[\Psi r^\ell Y_\ell^m(\theta,\phi)e^{-i\omega t}\right],
\end{equation}
where $(r,\theta,\phi)$ are spherical polar coordinates in an inertial
frame of reference centered on the star, $\omega$ is the frequency in
that frame, and $\Psi$ is an amplitude.  The tidal frequency
experienced by the star is
\begin{equation}
  \hat\omega=\omega-m\Omega,
\end{equation}
where $\Omega$ is its spin frequency.\footnote{We restrict our
attention to uniformly rotating stars.  By ``frequency'' we always
mean angular frequency.}  $Q'$ is a function of $\ell$, $m$, and
$\hat\omega$, but can be assumed to be independent of $\Psi$ if linear
theory applies.  In practice $\ell=2$ is dominant, and the azimuthal
wavenumber can be restricted to $m=0$, $1$, or $2$.

Although the potential applications of tides are very broad, we are
concerned here with two problems connected with observations: the
circularization of close binary stars and the inward migration of hot
Jupiters.

In a binary star with mean motion $n>0$, tidal dissipation leads to an
evolution of the semimajor axis $a$ and eccentricity $e$ at the rates
\begin{eqnarray}
  \f{\dot a}{a}&=&-\f{9}{2}\f{M_2}{M_1}\left(\f{R_1}{a}\right)^5n\left[\f{\sgn(2n-2\Omega_1)}{Q'_{1,2,2n-2\Omega_1}}\right]\nonumber\\
  &&-\f{9}{2}\f{M_1}{M_2}\left(\f{R_2}{a}\right)^5n\left[\f{\sgn(2n-2\Omega_2)}{Q'_{2,2,2n-2\Omega_2}}\right],
\label{adot}
\end{eqnarray}
\begin{eqnarray}
  \f{\dot e}{e}&=&-\f{9}{32}\f{M_2}{M_1}\left(\f{R_1}{a}\right)^5n
  \left[\f{49\,\sgn(3n-2\Omega_1)}{Q'_{1,2,3n-2\Omega_1}}\right.\nonumber\\
  &&\quad\left.+\f{6}{Q'_{1,0,n}}+\f{4\,\sgn(2\Omega_1-2n)}{Q'_{1,2,2n-2\Omega_1}}+\f{\sgn(2\Omega_1-n)}{Q'_{1,2,n-2\Omega_1}}\right]\nonumber\\
  &&-\f{9}{32}\f{M_1}{M_2}\left(\f{R_2}{a}\right)^5n
  \left[\f{49\,\sgn(3n-2\Omega_2)}{Q'_{2,2,3n-2\Omega_2}}\right.\nonumber\\
  &&\quad\left.+\f{6}{Q'_{2,0,n}}+\f{4\,\sgn(2\Omega_2-2n)}{Q'_{2,2,2n-2\Omega_2}}+\f{\sgn(2\Omega_2-n)}{Q'_{2,2,n-2\Omega_2}}\right],\nonumber\\
\label{edot}
\end{eqnarray}
where we neglect fractional corrections of second order in
eccentricity and obliquity.\footnote{Under the assumption that
$1/Q'\propto|\hat\omega|$, equation~(\ref{edot}) reproduces the result of
\citet{D80} that eccentricity is excited for $\Omega/n>18/11$ and
damped otherwise.  Note that the discontinuities in
equations~(\ref{adot}) and~(\ref{edot}) do not occur in practice
because $1/Q'$ must vanish as $\hat\omega$ tends to zero in the case
of a uniformly rotating body.}  Here $M$ and $R$ denote stellar mass and
radius, and $Q'_{i,m,\hat\omega}$ is the value of $Q'$ of star~$i$ for
$\ell=2$, azimuthal wavenumber~$m$, and tidal
frequency~$\hat\omega$.  For two identical stars that are
already synchronized ($\Omega_1=\Omega_2=n$), eccentricity is damped
on the circularization timescale
\begin{eqnarray}
  \tau_e&=&-\f{e}{\dot e}=\f{2}{63}\left(\f{a}{R}\right)^5\f{Q'}{n}\nonumber\\
  &\approx&1200\,Q'\left(\f{\bar\rho}{\bar\rho_\odot}\right)^{5/3}
  \left(\f{P}{10\,\mathrm{d}}\right)^{13/3}\,\mathrm{yr},
\label{te}
\end{eqnarray}
where $P$ is the orbital period and $\bar\rho$ is the mean density
($\bar\rho_\odot\approx1.41\,\mathrm{g}\,\mathrm{cm}^{-3}$).  The
timescale for synchronization is indeed much shorter.  Equation~(\ref{te})
assumes a common value of $Q'$ for the various tidal components, but
this assumption may not be so important because the
$\hat\omega=3n-2\Omega$ component dominates unless it has a much
larger value of $Q'$.

The relative contribution of the tide in star~2 is
$(Q'_1/Q'_2)(M_1/M_2)^2(R_2/R_1)^5$.  For late-type stars $R\propto
M^{4/5}$ approximately, giving a ratio of $(Q'_1/Q'_2)(M_2/M_1)^2$.
The tide in the secondary star is therefore less important if its $Q'$
is the same.

\begin{figure}
  \plotone{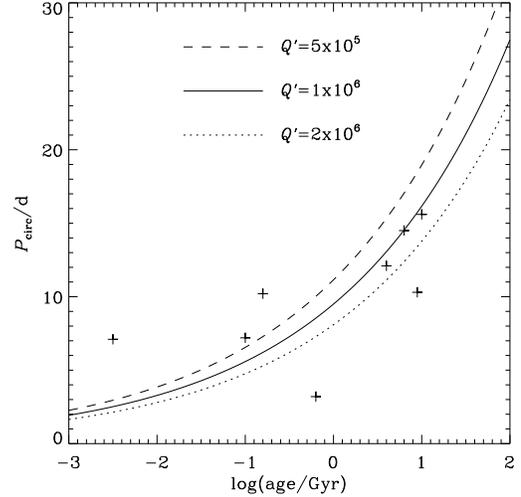}
  \caption{Circularization period versus age for the stellar clusters
  and populations listed in \citet{MM05}.  The curves show the period
  for which the circularization timescale~(\ref{te}) is equal to the
  age for various values of $Q'$.}
\label{f:meibom}
\end{figure}

Fig.~\ref{f:meibom} reproduces the observational data from the study
of \citet{MM05}.  Although the trend is not particularly clear it is
roughly consistent with the frequently quoted value $Q'\approx10^6$.

To determine the rate of inward migration of a hot Jupiter, we
consider a synchronized planet of mass $M_p$ in a circular orbit
around a star of mass $M_*$, in which case
\begin{equation}
  \f{\dot a}{a}=-\f{9}{2}\f{M_p}{M_*}\left(\f{R_*}{a}\right)^5n\left[\f{\sgn(2n-2\Omega_*)}{Q'_{*,2,2n-2\Omega_*}}\right].
\end{equation}
The inspiral time for a planet into a star similar to the Sun (assuming $Q'_*$ to be independent of frequency) is
\begin{eqnarray}
  \tau_a&=&-\f{2}{13}\f{a}{\dot a}=\f{4}{121}\f{M_*}{M_p}\left(\f{a}{R_*}\right)^5\f{Q'_*}{n}\nonumber\\
  &\approx&0.020\,Q'_*
  \f{M_*}{M_p}\left(\f{\bar\rho}{\bar\rho_\odot}\right)^{5/3}\left(\f{P}{\mathrm{d}}\right)^{13/3}\,\mathrm{yr}.
\label{ta}
\end{eqnarray}

The extrasolar planet of shortest orbital period found to date is \tr\
with $P=1.21\,\mathrm{d}$.  It orbits a star of mass $1.04\,M_\odot$
whose age is estimated as $3\pm1\,\mathrm{Gyr}$ \citep{S03}.  If the
value $Q'\approx10^6$ appropriate for binary stars is employed in this
context, the inspiral time is only about $0.036\,\mathrm{Gyr}$.

The observations raise two basic problems for theorists.  First, an
efficient dissipation mechanism must be found to explain a $Q'$ as
small as $10^6$ for the binary circularization problem.  Second, the
same mechanism must not operate so efficiently in the hot Jupiter
problem, unless we are extremely fortunate to observe \tr\ and others
like it.

\subsection{Equilibrium and dynamical tides}

The analysis of tides in stars was developed principally by
\citet{Z66a,Z66b,Z66c,Z70,Z75,Z77}.  In the theory of the equilibrium
tide, the tidal potential induces a time-dependent but
quasi-hydrostatic bulge.  Owing to the slow oscillation of the bulge a
certain velocity field is required within the star.  Dissipation
occurs in convective zones because an effective viscosity associated
with turbulent convection acts on this velocity field.  The resulting
value of $1/Q'$ is proportional to the effective viscosity, which
ought to be reduced when the tidal period exceeds the convective
timescale because then only part of the spectrum of turbulent
fluctuations damps the tide.

The assumption of a quasi-hydrostatic bulge appears appropriate
because the tidal frequency $|\hat\omega|$ is typically much smaller
than the dynamical frequency $\omega_*=(GM/R^3)^{1/2}$ of the star,
which means that fundamental and acoustic modes are driven well below
their natural frequencies.  However, stars support other types of wave of
much lower frequency.  The theory of the dynamical tide determines the
excitation of such waves by tidal forcing and their dissipation by
non-adiabatic processes, which can be effective if the wavelengths are
short.

Radiative zones are stably stratified and typically have a \bvf\ $N$
that is comparable to $\omega_*$ and much greater than $|\hat\omega|$.
Radial motions are strongly inhibited when $|\hat\omega|\ll N$.  On
the other hand $|\hat\omega|$ is typically comparable to $\Omega$, so
the Coriolis force can be important.  In uniformly rotating stars the
resulting solutions are called Hough waves and have a short radial
wavelength.  These reduce to gravity waves in non-rotating stars, as
considered in Zahn's work.

Convective zones are usually adiabatically stratified to high accuracy
and radial motions are not inhibited.  At tidal frequencies
$|\hat\omega|<2|\Omega|$ in uniformly rotating stars, convective zones
support pure inertial waves, for which the Coriolis force is the
restoring force.  These have an entirely different character from
Hough waves, although a subset of Hough waves are sometimes called
``inertial modes'' \citep[e.g.][]{SPA95}.  In a non-rotating star, or for
$2|\Omega|<|\hat\omega|\ll\omega_*$, convective zones do not support
waves.  Even in this case, as discussed below, the tidal response is
not described by the equilibrium tide.

\subsection{Previous theoretical approaches}

Several papers have previously presented calculations of the
efficiency of tidal dissipation in solar-type stars under a variety of
assumptions.  As for the theory of the equilibrium tide, some
controversy has surrounded the correct way to reduce the effective
viscosity when the tidal period exceeds the convective timescale, as
discussed by \citet{GO97}.  The theoretical arguments generally
support, at least in an approximate sense, the formula of \citet{GK77}
[cf.~equation~(\ref{nu_gk}) below], which drastically reduces the
effective viscosity at high frequency.  Recently, however,
\citet{PSRD07} find some support for the formula of \citet{Z66b} from an
analysis of numerical simulations of turbulent convection.  Even with
the more optimistic assumption of \citet{Z66b}, $Q'$ is too large to
explain binary circularization.

Both \citet{TPNL98} and \citet{GD98} extended the theory of the
dynamical tide, previously developed by Zahn for early-type stars with
convective cores, to solar-type stars with convective envelopes.  In
this approach, the tidal forcing excites an irrotational flow in the
convective zone and low-frequency gravity waves in the radiative
interior.  \citet{TPNL98} calculated adiabatic forced gravity waves,
which propagate through the entire radiative zone, and determined the
dissipation rate by including non-adiabatic effects as a perturbation.
Neither radiative damping nor turbulent viscosity is very effective in
damping the waves because they are evanescent in the convective zone.
The outcome was that tidal dissipation is efficient only when the
tidal frequency closely matches that of a global g~mode, but the
system would evolve very slowly between these resonances.  When the
dissipation rate is averaged in time, using an appropriately weighted
local average in frequency, the resulting value is dominated by the
behavior off resonance and corresponds to
$Q'\approx5.1\times10^7(P_\mathrm{tide}/10\,\mathrm{d})^{-0.15}$
[based on their equation~(37)] where
$P_\mathrm{tide}=2\pi/|\hat\omega|$ is the tidal period.  They used an
effective viscosity of the form recommended by \citet{GK77} but which
is in fact larger by a factor of $4\pi^2$ in the high-frequency limit
and closer to that used by \citet{GN77}.\footnote{Such factors are
certainly debatable as neither of the cited papers goes beyond simple
dimensional estimates.  Neither includes the factor of $1/3$ appearing
in equation~(\ref{nu_gk}) and used by subsequent authors.}  This value
of $Q'$ is substantially larger than that calculated on the basis of
the equilibrium tide,
$Q'\approx1.2\times10^7(P_\mathrm{tide}/10\,\mathrm{d})^{0.08}$ [based
on their equation~(36)], as that approximation gives an inaccurate
description of the disturbance in the convective zone.

\citet{GD98} did not calculate global g~modes but considered the
possibility that the gravity waves would reflect imperfectly from the
center of the star, where the wavelength is very short and
nonlinearity can occur because the wave energy is concentrated into an
extremely small volume.  Imperfect reflection could mean either that
the wave reflects from the center with a perturbed phase, or that the
wave is substantially dissipated at that location.  In either case
global g-mode resonances do not occur and tidal dissipation occurs
over a continuous range of tidal frequencies.  The result is
$Q'\approx1.2\times10^9(P_\mathrm{tide}/10\,\mathrm{d})^{8/3}$ [based
on their equation~(13)].  (When they converted this result into a
circularization timescale they apparently underestimated it by a
factor of about~20.)  The stronger frequency dependence in this
expression means that this mechanism could be very important for
short-period systems.

Neither of these studies included the effect of the Coriolis force on
the tidal disturbance, which might be very important if the tidal
frequency is not much larger than the spin frequency of the star.
Incorporating the Coriolis force is difficult because it leads to wave
equations that are not separable in the spherical polar coordinates
$r$ and $\theta$.  This separability is very useful because the
wavelength of the gravity waves can be so short that a two-dimensional
calculation with adequate resolution is unfeasible.  (The third,
azimuthal, direction can always be treated by a separation of
variables.)  \citet{SW02} and \citet{WS02} were able to retain
separability by using the ``traditional approximation'' in which the
latitudinal component $-\Omega\sin\theta$ of the angular velocity is
neglected in calculating the Coriolis force.  Indeed, the traditional
approximation is justified in the radiative region because radial
motions are suppressed for $|\hat\omega|\ll N$ and the relevant
solutions are Hough waves.  \citet{SW02} and \citet{WS02} developed a
highly sophisticated model including the effects of stellar evolution,
magnetic braking, and resonance locking.  The results of \citet{SW02}
are similar to those of \citet{TPNL98} off resonance, as they adopt
the same effective viscosity, but inclusion of the Coriolis force
allows a richer spectrum of resonances with various types of global
Hough modes.

\subsection{Purpose of this calculation}

Unfortunately the traditional approximation is not valid in the
convective zone, which is adiabatically stratified to high accuracy.
In a previous paper \citep[][hereafter Paper~I]{OL04} we studied the
excitation and dissipation of tidal disturbances in rotating giant
planets, which are mostly or fully convective.  As noted above, for
tidal frequencies $|\hat\omega|<2|\Omega|$, the convective zone
supports pure inertial waves, which are not correctly described by the
traditional approximation.  The importance of inertial waves for tidal
dissipation in giant planets has also been recognized by
\citet{W05a,W05b} and \citet{PI05}.

Inertial waves have remarkable properties, some of which have been
elucidated only since the advent of high-resolution numerical
calculations.  They propagate along characteristic rays that are
inclined to the rotation axis at a certain angle, depending on the
wave frequency.  This angle is necessarily preserved in reflections of
the waves from boundaries, so that a beam is typically either focused
or defocused in such a reflection.  When propagating within a
spherical annulus (rather than a full sphere), the waves are
generically focused on to ``wave attractors'' where intense
dissipation occurs \citep{RV97,RGV01}.  It was demonstrated
mathematically by \citet{O05} that the dissipation rate associated
with wave attractors in a slightly simplified wave equation is
independent of viscosity in the limit of very small Ekman number.  In
Paper~I we found evidence for a similar behavior in the limit of small
viscosity, although the numerical solutions indicate that the
dissipation is typically concentrated along the rays that emanate from
the critical latitude on the inner boundary.

The purpose of the present calculation is to determine the excitation
of inertial waves in the convective zones of solar-type stars and to
assess their role in tidal dissipation.  This occurs both through the
damping of the waves in situ by turbulent viscosity and also through
their effect on the excitation of Hough waves in the radiative zone.
We aim to determine what range of values of $Q'$ can be obtained for
the binary circularization problem and the hot Jupiter problem.

In some sense our calculation is complementary to that of
\citet{SW02}.  They gave an accurate treatment of the radiative zone
but not the convective zone, and assumed that Hough waves reflect
coherently from the center of the star.  We aim to treat the
convective zone more accurately, and assume that the Hough waves do
not reflect coherently from the center.

\section{Numerical analysis}

\subsection{Stellar model}

For an accurate and detailed model of the present Sun, we adopt
``model S'' from \citet{CD96}.  In the mixing-length theory of
\citet{BV58}, the convective energy flux density is related to the
convective velocity $v$ by
\begin{equation}
  F_c=\f{1}{3}\f{8v^3}{\ell}\f{\rho c_pT}{g\delta}
\end{equation}
where $\delta=-(\p\ln\rho/\p\ln T)_p$ and $\ell=\alpha H_p$ is the
mixing length, $H_p=p/\rho g$ being the pressure scaleheight.  [See
\citet{Z89} for a discussion of the dimensionless numerical prefactors.]
The convective timescale is then $\tau=\ell/v$.

The turbulent viscosity based on mixing-length theory is
${\textstyle\f{1}{3}}v\ell$, but this should be reduced when the tidal
period exceeds the convective timescale.  We take the effective
turbulent viscosity to be
\begin{equation}
  \nu=\f{1}{3}v\ell\left(1+\hat\omega^2\tau^2\right)^{-1}
\label{nu_gk}
\end{equation}
which is a smoothed version of that of \citet{GK77}.  In fact, the
functional form of equation~(\ref{nu_gk}) is suggested by a Maxwellian
viscoelastic model in which $\tau$ is identified with the relaxation
time of the turbulent stress.

\begin{figure}
  \plotone{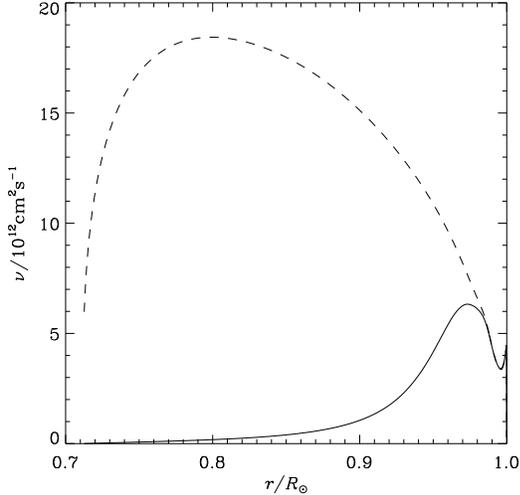}
  \caption{Effective viscosity profile in the convective zone of the
  present Sun.  The mixing-length value
  $\nu={\textstyle\f{1}{3}}v\ell$ is shown as a dashed line, while the
  value reduced according to equation~(\ref{nu_gk}) for a tidal
  period of 10~d is shown as a solid line.  For a spin period of 10~d
  a kinematic viscosity of
  $\nu=10^{12}\,\mathrm{cm}^2\,\mathrm{s}^{-1}$ corresponds to an
  Ekman number of $Ek=\nu/2\Omega R^2\approx1.4\times10^{-5}$.}
\label{f:nu}
\end{figure}

Fig.~\ref{f:nu} shows the viscosity profile in the Sun for a tidal
period of 10~d.  This period is chosen as being representative of the
binary circularization problem.  The viscosity is substantially
reduced below the mixing-length value in most of the convective zone.
For reasonable spin periods, the dimensionless Ekman number
$\nu/2\Omega R^2$ is small but not extremely small (typically of the
order of $10^{-5}-10^{-4}$).  This means that inertial waves will be
modestly affected by viscosity, and the problem is conveniently
accessible to computation with achievable numerical resolution.

Although the interior angular velocity profile of the Sun has been
measured, we assume the star to be uniformly rotating.  This is partly
for technical reasons, as our numerical method can cope easily only
with ``shellular'' rotation profiles.  In any case it is not clear how
the rotation profile would change if the Sun were tidally synchronized
at a different rate in a binary system.

\subsection{Inertial waves in the convective zone}

We solve the forced linearized equations in the convective zone as in
Paper~I.  The response consists of an equilibrium tide plus a
dynamical tide.  The former satisfies a second-order ordinary
differential equation in $r$ which we solve numerically.  The latter
satisfies partial differential equations in $r$ and $\theta$ which can
be simplified in the low-frequency limit, the simplifications
amounting to the Cowling and anelastic approximations.  The velocity
field is separated into spheroidal and toroidal parts and the
equations are converted into a large algebraic system by Galerkin
projection on to normalized associated Legendre polynomials
(i.e.~spherical harmonics) in $\theta$ and Chebyshev collocation in
$r$.  The linear system is solved by a standard method for block
tridiagonal matrices, and the total viscous dissipation rate is
calculated.  We adopt stress-free impermeable boundary conditions on
the dynamical tide at the upper and lower limits of the convective
zone, as explained in Paper~I.

\subsection{Hough waves in the radiative zone}

We determine algebraically the excitation of Hough waves in the
radiative zone as in Paper~I.  They are excited at the interface
between the radiative and convective zones, partly by tidal forcing in
that region and partly by the pressure of the inertial waves acting at
the interface.  Following \citet{GD98}, the model assumes that $N^2$
vanishes in the convective zone and increases initially linearly with
distance into the radiative zone.  We assume that the waves are not
reflected coherently from the center of the star and calculate the
resulting energy flux.  This is converted into a dissipation rate as
described in Paper~I.  The behavior of waves near the center of the
star in the limit $|\hat\omega|\gg|\Omega|$ is discussed in
Appendix~\ref{s:center}, where, in a slight refinement of the
calculation of \citet{GD98}, we estimate the conditions under which
the waves become nonlinear.

\section{Results}
\label{s:results}

\subsection{Inertial waves}

\begin{figure*}
  \plottwo{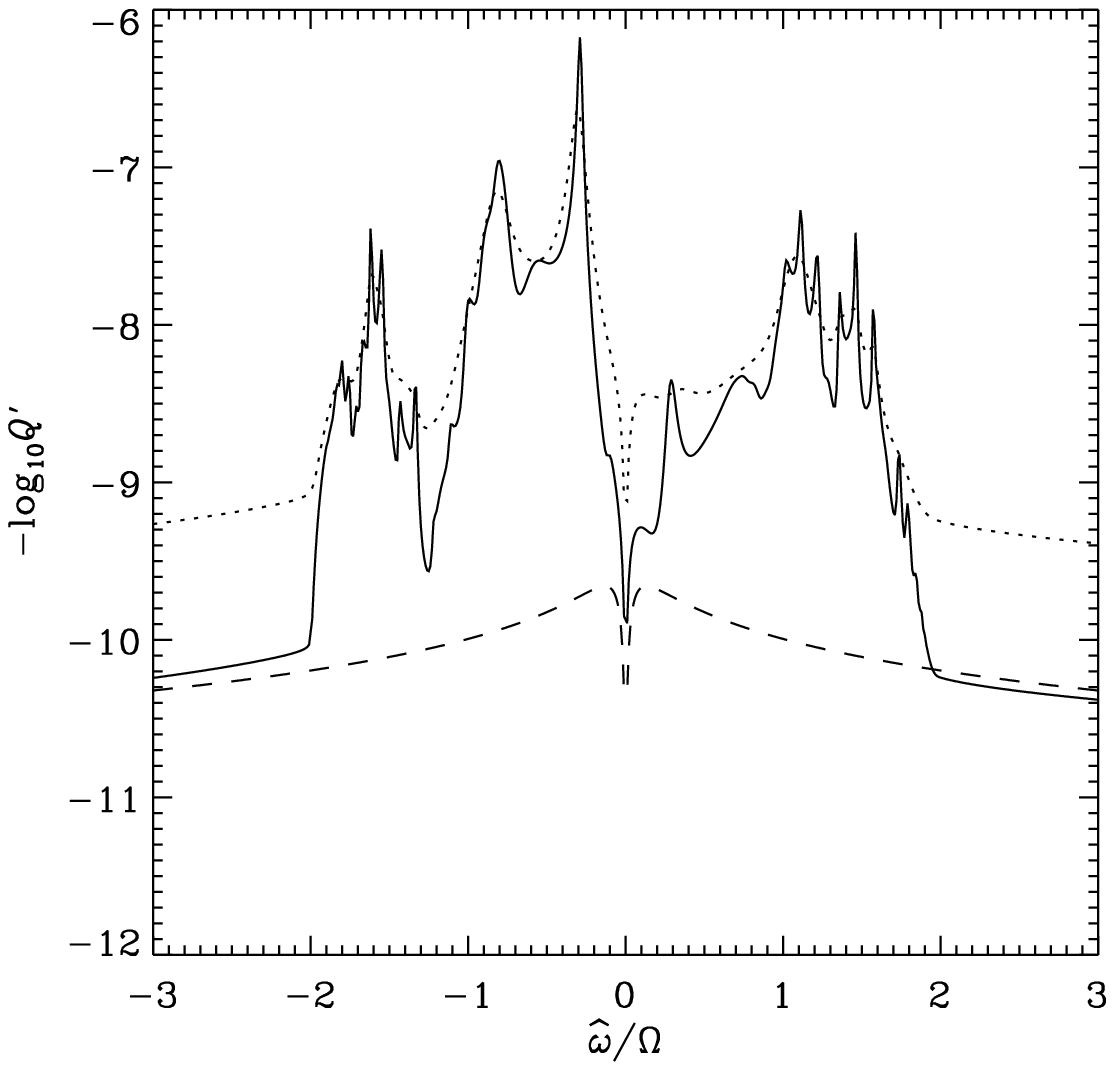}{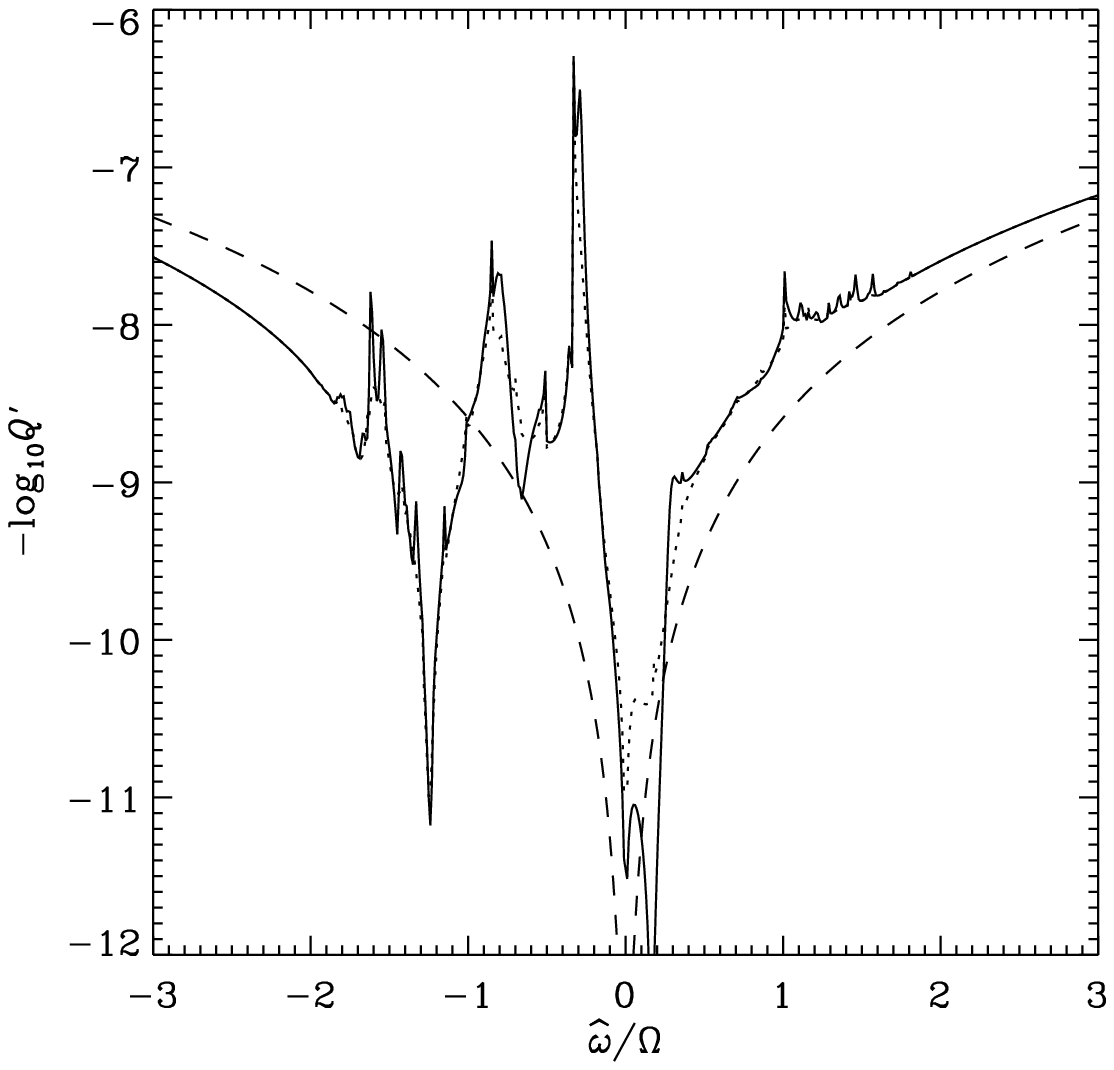}
  \caption{Dissipation rate, parametrized as a value of $Q'$, as a
  function of tidal frequency.  These results refer to tidal forcing
  by the $\ell=m=2$ solid harmonic in a model of the Sun with a spin
  period of 10~d.  \textit{Left}: $Q'$ from the viscous dissipation of
  inertial waves in the convective zone.  \textit{Right}: $Q'$ from
  the excitation of Hough waves in the radiative zone.  The dashed
  lines show the effect of omitting the Coriolis force.  In this
  figure only, the dotted lines show the result of artificially
  increasing the effective viscosity by a factor of~10.  }
\label{f:1msun_10d}
\end{figure*}

In the left half of Fig.~\ref{f:1msun_10d} we plot, as a function of
the tidal frequency, the value of $Q'$ resulting from the viscous
dissipation of inertial waves in the convective zone of the Sun with a
spin period of 10~d.  Only the most important azimuthal wavenumber
$m=2$ is considered.  Also shown for comparison is the viscous
dissipation rate of the irrotational disturbance generated in the
convective zone when the Coriolis force is omitted.  For frequencies
$|\hat\omega|<2|\Omega|$ the dissipation is greatly enhanced by the
excitation of inertial waves, but outside this range the Coriolis
force has little net effect.

Where the dissipation rate is significantly enhanced, it is relatively
insensitive to the viscosity, as found in Paper~I.  Elsewhere the
dissipation rate is directly proportional to the viscosity.  The value
of $Q'$ obtained when the Coriolis force is omitted is larger than
that found by \citet{TPNL98}.  This is partly because of the
contribution of g-mode resonances to their average value, but mainly
because their viscosity is larger than ours by a factor of
approximately $4\pi^2$.

The numerical convergence of these results was verified by repeating
the calculation with double the resolution in each direction.  It was
found adequate in most cases to truncate the Legendre and Chebyshev
polynomial bases at an order of~100.

\subsection{Hough waves}

Also shown in Fig.~\ref{f:1msun_10d} is the same information for Hough
waves excited at the interface between the radiative and convective
zones.  Inclusion of the Coriolis force can either increase or
decrease the dissipation rate.

When the Coriolis force is neglected, our results should agree with
those of \citet{GD98}.  We find their numerical parameter $\sigma_c$
to equal $-1.19$ rather than $-0.64$, which we verified by numerically
integrating their equation~(3).  Altogether we find
$Q'\approx3.9\times10^8(P_\mathrm{tide}/10\,\mathrm{d})^{8/3}$ when
spin is neglected.  We also obtain their equation~(15) but with
$(M_1+M_2)/M_1$ raised to the power $-5/3$ instead of $+11/6$.  This
discrepancy suggests that they may have overestimated the
circularization rate by a factor of $2^{7/2}$.

\subsection{Higher tidal frequencies}

\begin{figure*}
  \plottwo{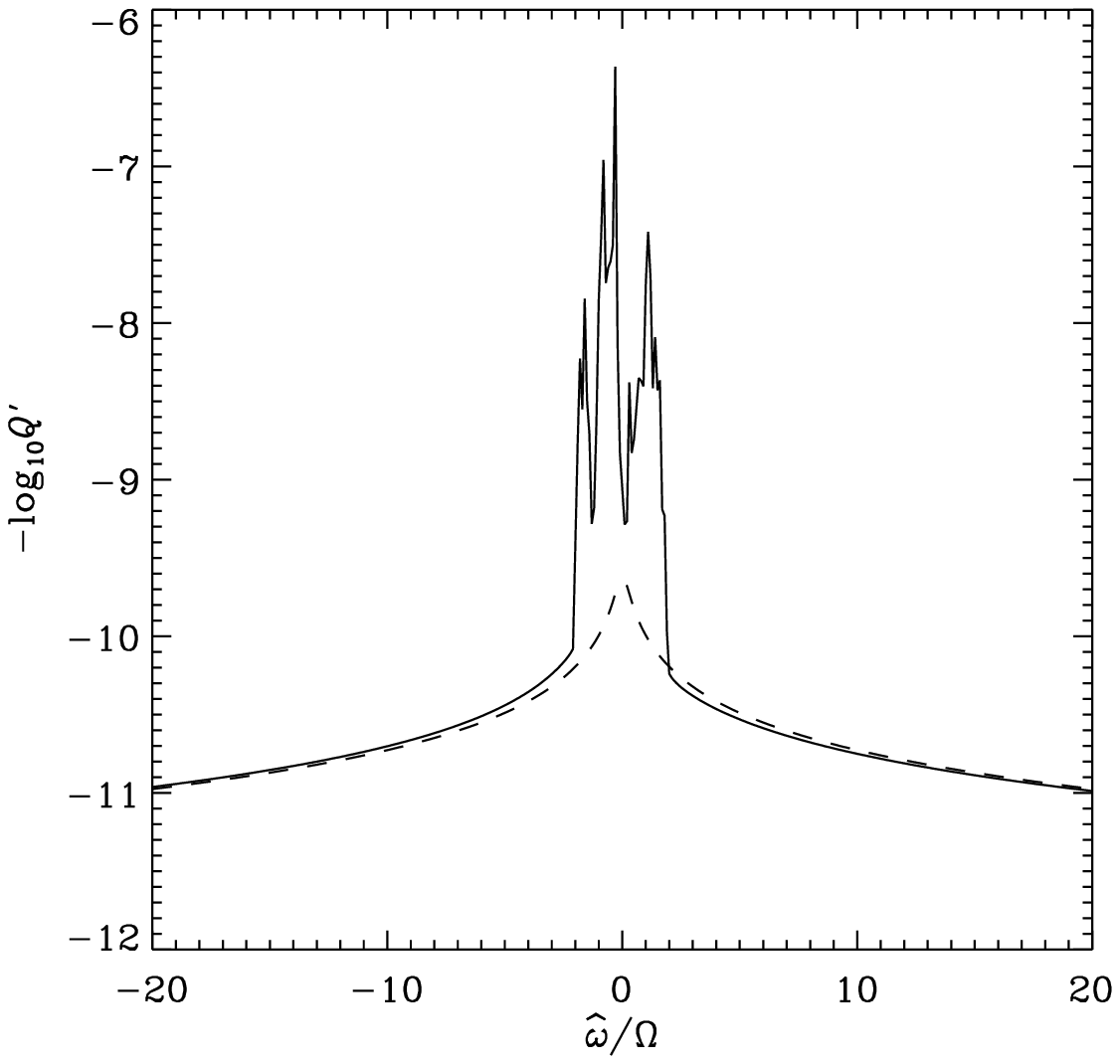}{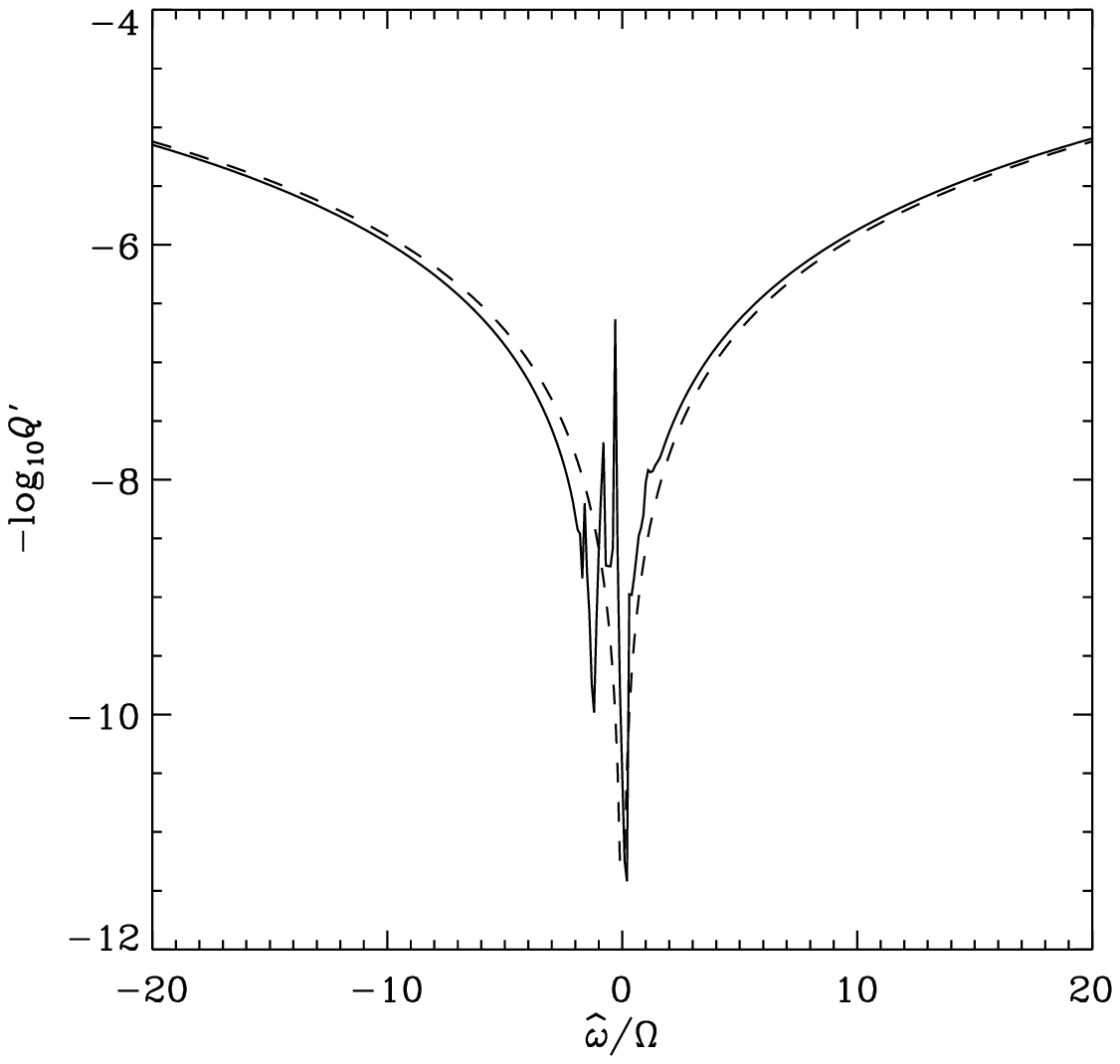}
  \caption{Expansion of Fig.~\ref{f:1msun_10d} to a wider range of
  tidal frequencies.}
\label{f:1msun_10d_long}
\end{figure*}

In Fig.~\ref{f:1msun_10d_long} we plot similar results over a wider
range of tidal frequencies.  As expected, the Coriolis force is
unimportant when $|\hat\omega|\gg|\Omega|$.  As found by \citet{GD98},
Hough waves (or gravity waves, in this regime) provide efficient tidal
dissipation at high frequencies if they do not reflect coherently from
the center of the star.

\subsection{Dependence on spin period}

\begin{figure*}
  \plottwo{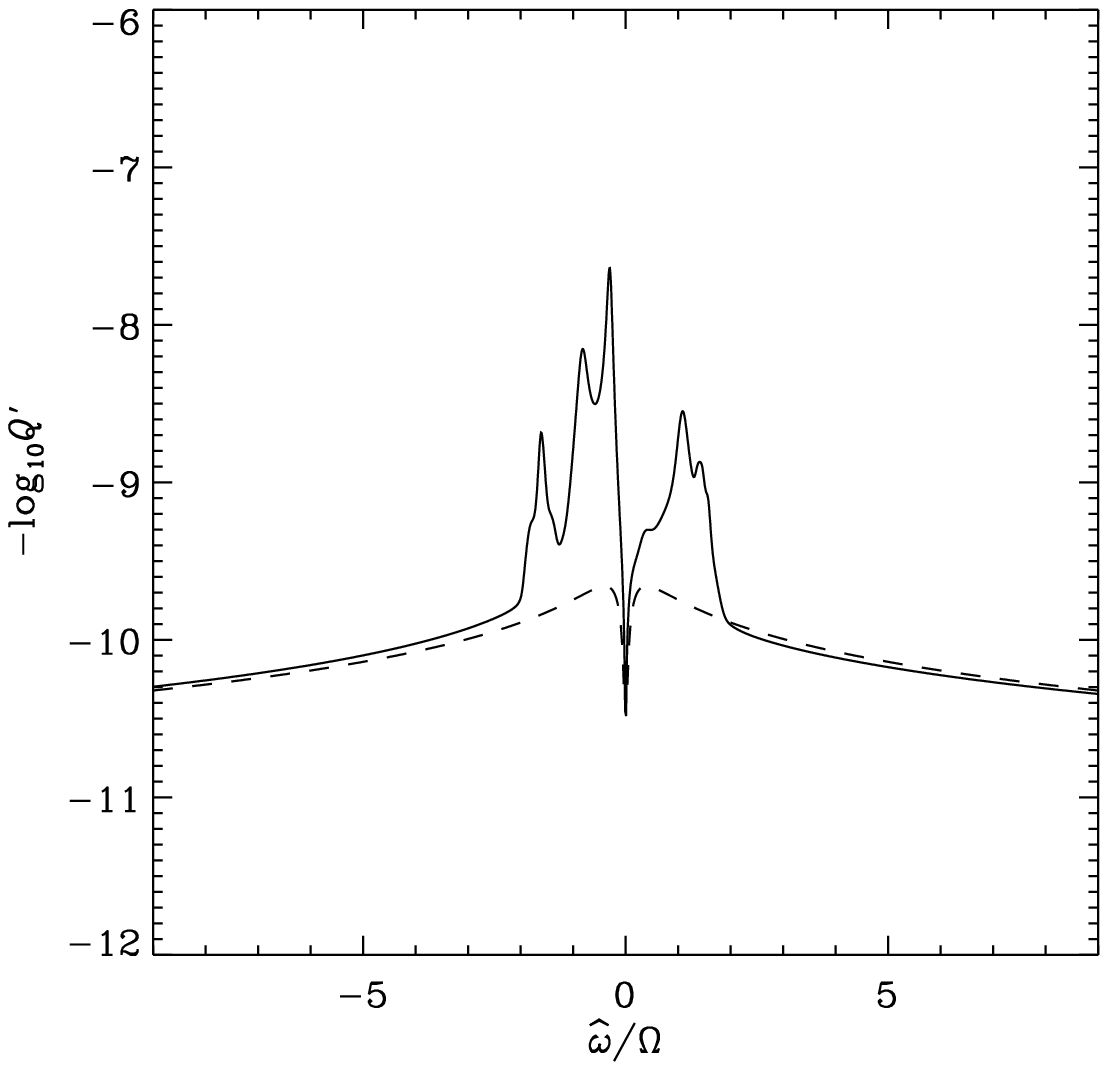}{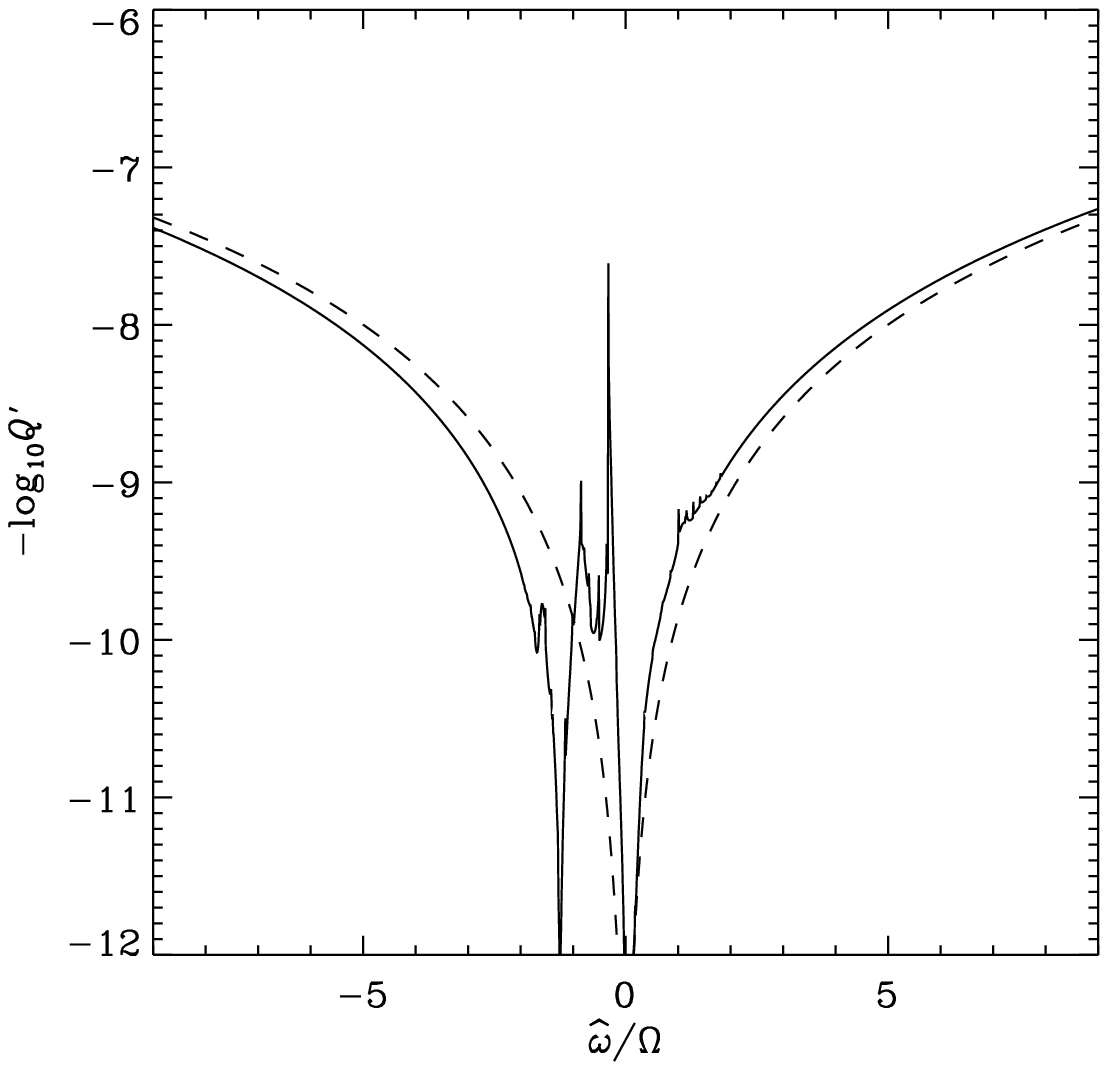}
  \caption{As for Fig.~\ref{f:1msun_10d}, but for a spin
  period of 30~d.  Note the wider range of $\hat\omega/\Omega$,
  corresponding to the same range of $\hat\omega$ as in
  Fig.~\ref{f:1msun_10d}.}
\label{f:1msun_30d}
\end{figure*}

\begin{figure*}
  \plottwo{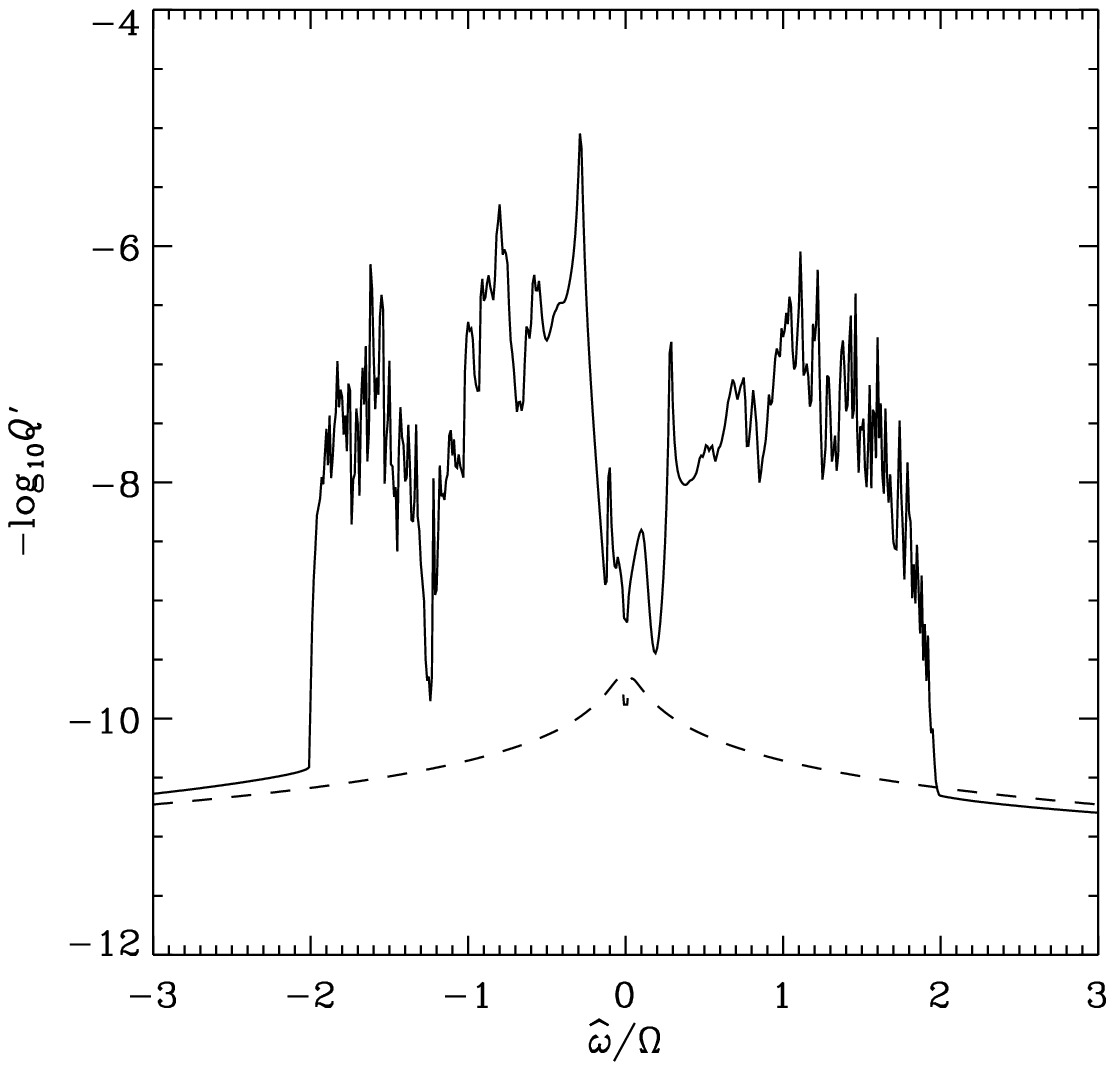}{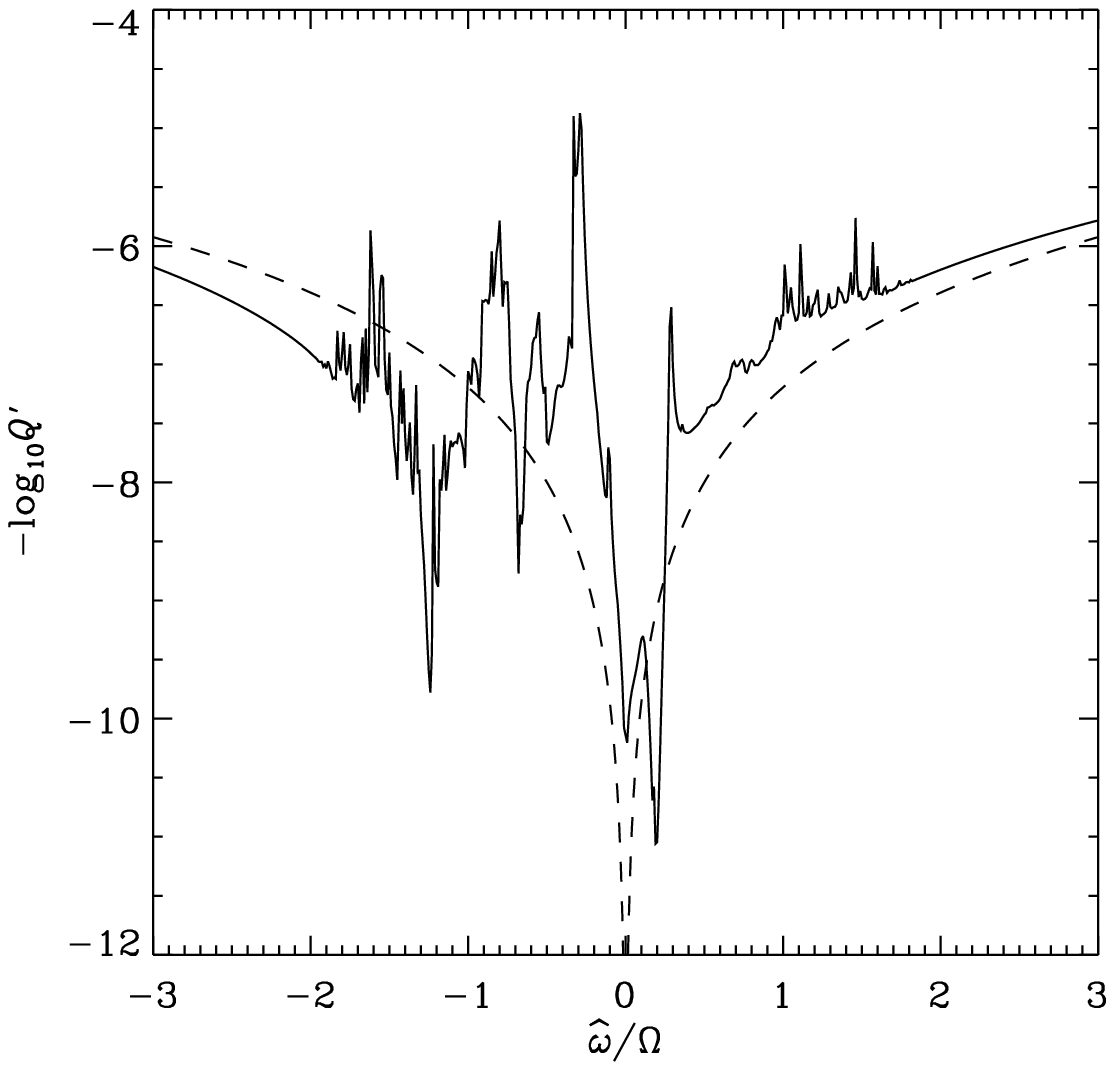}
  \caption{As for Fig.~\ref{f:1msun_10d}, but for a spin
  period of 3~d.  Note the different vertical scale.}
\label{f:1msun_3d}
\end{figure*}

In Fig.~\ref{f:1msun_30d} we present the results for the Sun with a
spin period of 30~d.  Three times the range of $\hat\omega/\Omega$ is
plotted, corresponding to the same range of $\hat\omega$ as in
Fig.~\ref{f:1msun_10d}.  As expected, outside the range
$|\hat\omega|<2|\Omega|$, $Q'$ is almost independent of the spin
frequency $\Omega$ at a given tidal frequency $\hat\omega$, because
the effect of the Coriolis force is weak.  Within the frequency range
of inertial waves, a smaller $Q'$ is generally achieved in the more
rapidly rotating star.  Fig.~\ref{f:1msun_3d} shows the results for a
spin period of 3~d.  An even smaller $Q'$ results from inertial waves
in this case.  The scaling is approximately $1/Q'\propto\Omega^2$ for
a fixed value of $\hat\omega/\Omega$, as suggested by the factor
$f_Q$ defined in Paper~I.  Increasing $\Omega$ at fixed
$\hat\omega/\Omega$ also decreases the Ekman number, resulting in a
more structured graph as the system is less controlled by viscosity.

\subsection{Lower-mass stars}

We also investigated models of lower mass, as the secondaries or even
the primaries of ``solar-type binaries'' may have masses significantly
less than $1\,M_\odot$.  The greater depth and density of the
convective envelopes of these stars allows for a more efficient
excitation of inertial waves, in the sense that the dimensionless
dissipation rate $D_\mathrm{visc}/U_\mathrm{visc}$ (as defined in
Paper~I) is generally larger.  However, this effect is offset by the
fact that the conversion factor $f_Q$ for these stars is smaller as a
result of their greater mean density.

\begin{figure*}
  \plottwo{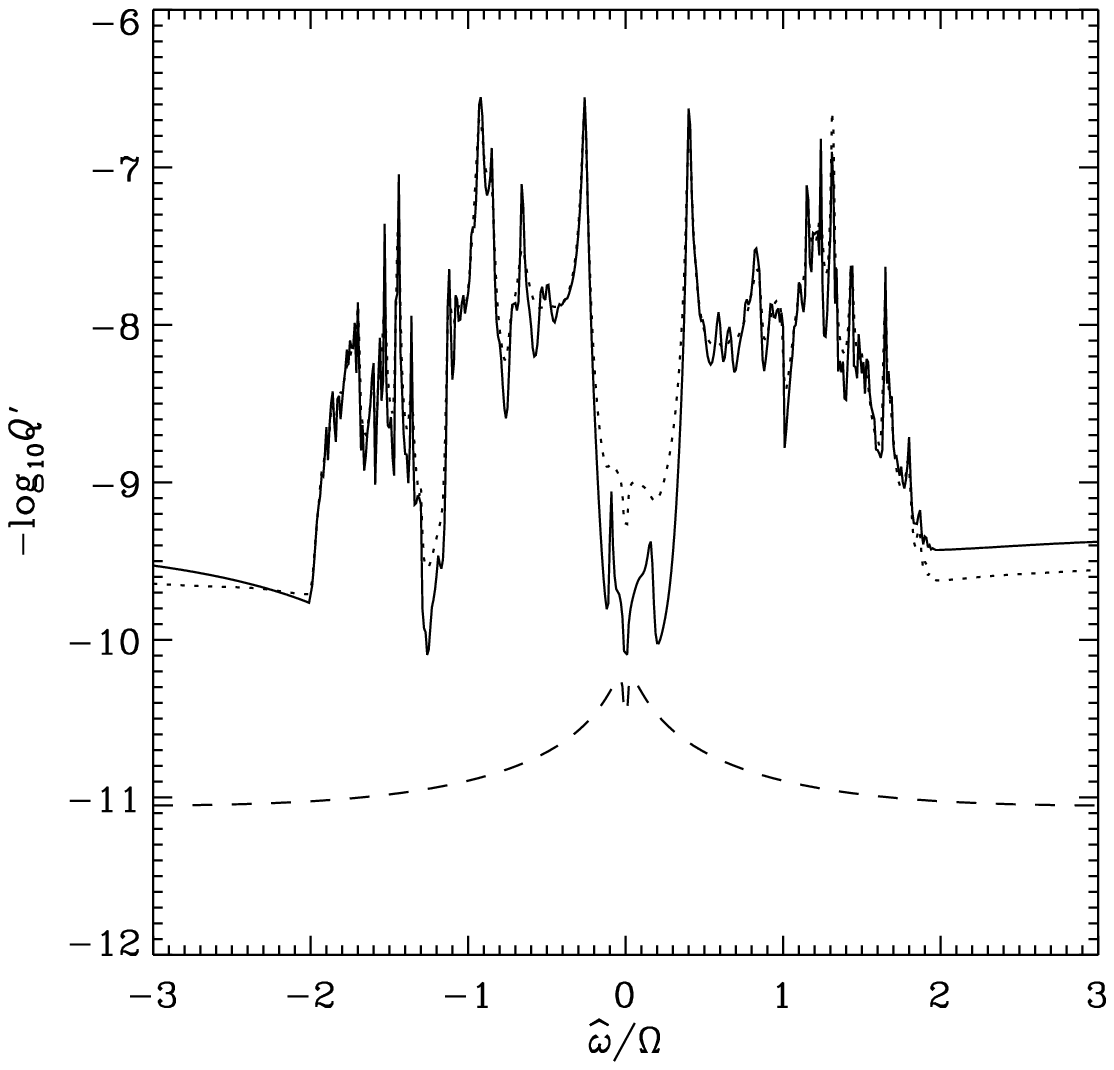}{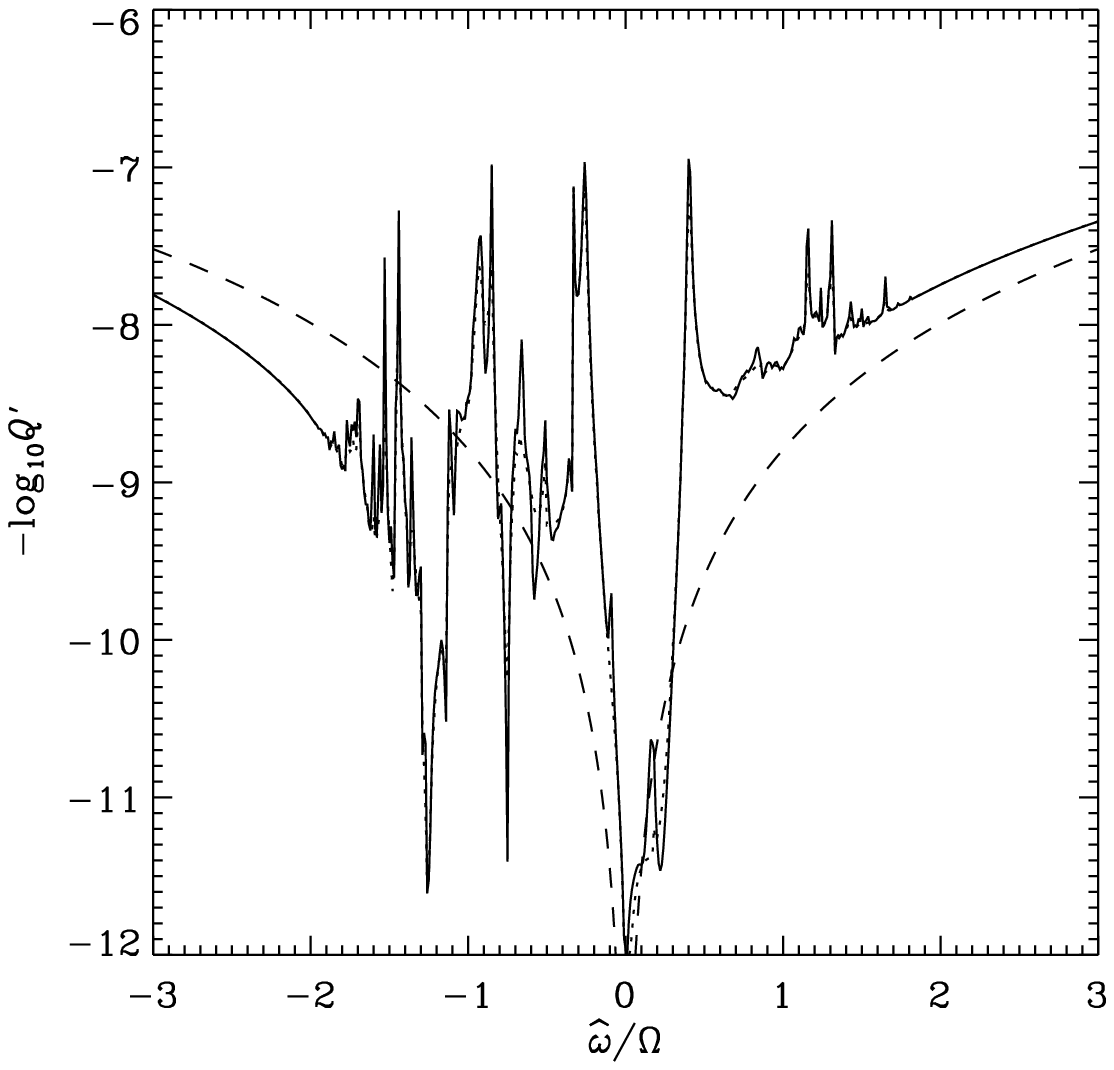}
  \caption{As for Fig.~\ref{f:1msun_10d}, but for a star of
  mass $0.5\,M_\odot$ and age 5~Gyr.}
\label{f:0.5msun_10d}
\end{figure*}

In Fig.~\ref{f:0.5msun_10d} we plot the values of $Q'$ resulting from
inertial and Hough waves in a stellar model of mass $0.5\,M_\odot$,
age $5\,\mathrm{Gyr}$, and spin period 10~d.  Although the details are
different, the qualitative behavior and the range of values of $Q'$
obtained are similar to the case of the Sun.

\subsection{Summary}

In the frequency range $|\hat\omega|<2|\Omega|$, tidal dissipation in
the convective zones of solar-type stars is substantially enhanced
through the excitation of inertial waves when the full Coriolis force
is included.  Here $Q'$ can be decreased by up to four orders of
magnitude and has a complicated dependence on tidal frequency.
Typically $Q'\propto\Omega^{-2}$ for a fixed ratio
$\hat\omega/\Omega$.  Values as small as $10^6$ can be achieved if the
star spins more rapidly than the Sun.

If the Hough waves excited at the interface between the radiative and
convective zones do not reflect coherently from the center of the star
as a result of their nonlinearity, they provide another means of
dissipation at all frequencies.  The resulting value of $Q'$ scales
generally with $|\hat\omega|^{-8/3}$ although this is modified,
especially in the range $|\hat\omega|<2|\Omega|$, by the inclusion of
the Coriolis force.  Values as small as $10^6$ can be achieved if the
tidal period is as short as 1~d.  The estimates in
Appendix~\ref{s:center} suggest that, in the case of the present Sun,
Hough waves become nonlinear in eccentric binaries but probably not in
the hosts of hot Jupiters.  Nonlinearity is less likely in stars that
are younger or less massive than the Sun, but is more likely in older
stars.

\section{Comparison with observed systems}

\subsection{Tidally induced orbital migration}

\subsubsection{Close-in gas giant planets around G dwarfs}

One immediate application of the present analysis is to the orbital
migration of close-in extrasolar planets.  In his study of \tr,
\citet{S03} computed the orbital decay timescale $a/|\dot a|$ (i) by
extrapolating a formula calibrated on the circularization of binary
stars, (ii) from equilibrium tidal models calculated using the
prescription of \citet{Z66b} for the effective viscosity in the
convective envelope, and (iii) by a similar method using the
prescription of \citet{GK77}.  (He set the tidal period equal to the
orbital period, thereby underestimating the forcing frequency by a
factor of approximately~2.  The tidal frequency associated with the
appropriate $m=2$ potential component in the hot Jupiter problem is in
fact $\hat\omega=2n-2\Omega$.)  He showed that, while the
theoretically derived $a/|\dot a|$ exceeds 10~Gyr with
prescription~(ii) and 4000~Gyr with prescription~(iii), is it less
than $1\,\mathrm{Gyr}$ with prescription~(i).  This discrepancy is
similar to that found in equilibrium tidal models for solar-type
binary stars (see Section~\ref{s:solarbinary}).

To the extent that the mechanism and efficiency of tidal dissipation
remain uncertain, prescription~(i) should provide a useful empirical
estimate.  \citet{S03} also estimated the age of OGLE-TR-56 as
$\tau_*\approx3\pm1\,\mathrm{Gyr}$, which is considerably longer than
his estimate for $a/|\dot a|$ based on prescription~(i).  Note further
that, owing to the accelerating nature of inward orbital migration,
the inspiral time $\tau_a$ is considerably shorter than the present
value of $a/|\dot a|$ [see equation~(\ref{ta})].  While \tr\ was the
only known extrasolar planet with a period close to 1~d, a possible
interpretation was that it does indeed have $\tau_a\ll\tau_*$ and we
observe it with a small probability ($\sim\tau_a/\tau_*$) while it
evolves through this period range with a rapidly decreasing $\tau_a$
($\propto P^{13/3}$ under the assumption of constant $Q'$).  However,
two additional extrasolar planets, OGLE-TR-113b and OGLE-TR-132b, have
been found with only slightly longer orbital periods than \tr.  The
values of $\tau_a$ obtained for these planets under prescription~(i)
are also comparable to or shorter than their estimated ages.  The
period distribution of the known planets does not statistically
indicate that these very short-period planets are at the endpoints of
their tidal orbital evolution.  An alternative interpretation of these
data is that the relevant value of $Q'$ of the solar-type host stars
may be significantly larger than the one that applies to binary
circularization.

Based on the results presented in Section~\ref{s:results}, we explore
the possibility that this dichotomy in the stellar $Q'$ may be caused
by the differences in the tidal forcing frequencies and spin
frequencies for the hot Jupiter and binary circularization problems.
The shortest orbital periods of known extrasolar planets are little
longer than 1~d, which is an order of magnitude smaller than the
circularization periods of mature solar-type binary stars.  Owing to
selection effects, the spin periods of the known planet-bearing stars
are comparable to or longer than that of the Sun.  (The absence of
significant chromospheric activity, which is generally correlated with
rapid stellar spin, has been an important criterion in the selection
of target stars for radial-velocity surveys.)  In contrast,
circularized binary stars appear to attain spin synchronization so
that their spin period is roughly 10~d \citep{MMS06}.  This state is
maintained despite the stars' loss of spin angular momentum through
magnetic braking because this loss can be compensated by the tidal
transfer from the orbit in the case of two massive companions.
Extrasolar planets have much less orbital angular momentum and the
tidal interaction between the star and the planet cannot compensate
for the loss of spin angular momentum through magnetic braking.

The important difference is that the tidal forcing frequency falls
within the range of inertial waves in the binary circularization
problem and well outside it for the hot Jupiter problem.  In a
synchronized binary the eccentricity tides have frequencies
$\hat\omega=\pm\Omega$ (cf.~Section~\ref{s:timescales}) and can
efficiently excite inertial waves, leading to an enhanced tidal
dissipation.  In contrast, planetary companions in circular orbits are
unable to excite inertial waves in their host stars unless
$\Omega_*>n/2$.  In Fig.~\ref{f:1msun_10d_long} the tidal forcing by a
planet in a 1~d orbit falls at $\hat\omega/\Omega=18$ and results in
$Q'\approx8.9\times10^{10}$ from viscous dissipation in the convective
zone; this conclusion depends only weakly on the stellar spin for
$|\Omega|\ll n$.  On the other hand, Fig.~\ref{f:1msun_10d_long} shows
that Hough waves would be very important in the tidal evolution of
short-period planets, leading to $Q'\approx1.6\times10^5$ for
$P=1\,\textrm{d}$, provided that they fail to reflect coherently from
the center of the star.  The nonlinearity parameter is estimated in
Appendix~\ref{s:center} as $A\approx320(M_p/M_*)(P/\textrm{d})^{1/6}$
in the case of the present Sun, which is about $0.3$ for \tr\ and
somewhat less than unity for hot Jupiters generally.  The nonlinearity
increases with stellar mass and age.  This estimate shows how
important it is to determine the fate of Hough waves (or gravity
waves, in this regime) approaching the center for various levels of
nonlinearity.

In contrast, for binary stars in a nearly synchronous state in which
$\hat\omega\approx\Omega\approx n$ for the dominant eccentricity tide,
the Coriolis force strengthens the energy dissipation associated with
both inertial and Hough waves.  Over most of the frequency range
$|\hat\omega|<2|\Omega|$, the values of $Q'$ associated with inertial
waves are 2--3 orders of magnitude smaller than those outside the
range.  This frequency dependence makes it possible for the
dissipation of inertial waves to speed up the circularization of
binary stars (see Section~\ref{s:solarbinary}) without significantly
influencing the orbital evolution of the close-in planets.

Planets are formed in protostellar disks.  Classical T~Tauri stars are
observed to rotate with periods of a few days \citep{BCFMM93} and they
spin up as they contract on to the zero-age main sequence.  These
rapid spin rates are sustained for $\tau_\ast<100\,\textrm{Myr}$
\citep{BFA97}.  During this early phase of stellar evolution, the host
stars of the very short-period planets may have had $\Omega_*>n/2$,
allowing the excitation of inertial waves.  A relatively small $Q'$
can occur in this regime owing to the rapid spin (see
Fig.~\ref{f:1msun_3d}).
In addition, the properties of the convection may have been different.
Nevertheless, this phase may contribute little to orbital evolution as
it is short-lived.

\subsubsection{Close-in brown dwarf around an M dwarf}
\label{s:41004a}

A precision radial velocity survey also led to the discovery of an
$M_p\sin i=19\,M_J$ planet HD~41004Bb on a 1.33~d orbit around a
0.4~$M_\odot$ M-dwarf star \citep{ZMSUM03}.  This period is the
shortest known for any brown dwarf companion, and the estimated age of
the system is $\tau_\ast=1.6\pm0.8$ Gyr.  Assuming that the M dwarf
fails to spin synchronously because of the effects of magnetic
braking, we find from equation~(\ref{ta}) that $\tau_a\approx
\tau_\ast$ when $Q'\approx3.7\times10^7$.  Alternatively, if we apply
the frequency-independent $Q'$-value ($\sim 10^6$), calibrated on the
circularization of solar-type binary stars, the extrapolated value of
$\tau_a$ would be nearly two orders of magnitude smaller than the
estimated age.

With its low mass, the host star HD~41004B has a smaller radiative
core than the Sun.  Since it has no measurable rotation speed
\citep{B06}, the tide raised by the brown dwarf companion is likely to
be well outside the frequency range of inertial waves where $Q'$ is
relatively small.  The results in Section~\ref{s:results} indicate that
if the spin period of HD~41004B is longer than about 3~d, viscous
dissipation in the convective zone provides a $Q'$ much larger than
$10^7$ and poses no threat to the survival of the brown dwarf.
However, if during the early phases of HD~41004B's evolution $\Omega$
was substantially larger than its present value, the tidal frequency
could have fallen in the range of inertial waves.  Provided the
resulting low $Q'$ did not last more than $10^8$ yr, the amount of
orbital decay would still have been negligible.

Hough waves can also be excited in the radiative zone of HD~41004B.
Despite the relatively large mass ratio $M_p/M_*$ of this system, the
small value of $dN/dr$ at the center of the $0.4\,M_\odot$ star means
that the Hough waves are unlikely to become nonlinear (see
Appendix~\ref{s:center}).  This is important because if the Hough
waves were effectively dissipated they would lead to $Q'<10^6$ and
seriously threaten the survival of the brown dwarf.

\subsection{Tidal circularization}

\subsubsection{Solar-type binary stars}
\label{s:solarbinary}

In the introduction, we indicated that the dissipation of tidal
disturbances in both the convective and radiative zones of
non-rotating solar-type stars fails, by about two orders of magnitude,
to match that needed to account for the variation of the
circularization period of binary stars as a function of their age (see
Fig.~\ref{f:meibom}).  These discrepancies are particularly noteworthy
since the previously obtained values of $Q'$ might be underestimated
as a result of optimistic assumptions.  For example, in the
application of the \citet{GK77} formula, a factor of $4\pi^2$ was
introduced when \citet{TPNL98} replaced $\hat\omega\tau$ with
$\tau/P_\mathrm{tide}$.  The discrepancies would be still larger if
the prescription (\ref{nu_gk}) were applied.  Similarly, \citet{S03}
set the tidal frequency equal to the orbital frequency and thereby
underestimated the suppression of the turbulent viscosity.  Finally,
\citet{GD98} assumed in one section of their paper that Hough waves
are unable to reflect coherently from the center and form global
standing modes, thereby allowing effective dissipation at all
frequencies.

The results presented in this paper show that the Coriolis force can
introduce a frequency dependence in and reduce the magnitude of $Q'$.
The timescale for the stars to synchronize is generally much shorter
than that required for circularization.  The tidal frequency
$\hat\omega_c=3n-2\Omega$ of the principal tidal component involved in
circularization is larger than that involved in synchronization
($\hat\omega_s=2n-2\Omega$).  Eccentricity can be excited for
sufficiently large spin frequency $\Omega$ when $\hat\omega_c$ becomes
negative.  However, during the spin-down of the star towards
synchronization, $\hat\omega_s$ tends to zero, $\hat\omega_c$ becomes
positive and tends to $\Omega$, and the eccentricity of the binary is
damped.  For this tidal frequency, the results in
Fig.~\ref{f:1msun_10d} indicate that the values of $Q'$ associated
with inertial and Hough waves are $5.7\times10^7$ and $1.1\times10^8$,
respectively.  These values are an order of magnitude larger than that
obtained for equilibrium and dynamical tides in non-rotating stars if
the gravity waves reflect from the stellar center \citep{TPNL98}.  But
our values of $Q'$ are still an order of magnitude too large to
account for the observational data in Fig.~\ref{f:meibom}.  Thus the
incorporation of the Coriolis force reduces but does not eliminate the
discrepancies between theory and observations.

There are several potential mechanisms that may contribute to the
circularization process.
\citet{WS02} made an attempt to take the effects of stellar evolution
into account.  They suggested that binary systems may undergo
a complicated evolution through resonant tidal interactions.
In principle a strong tidal dissipation efficiency at
$\hat\omega_c\approx\Omega$ could lead to effective eccentricity
damping after the binary stars are synchronized.  But the protracted
maintenance of this feature is needed over timescales comparable to
$\tau_e$.

The calculations described in this paper are carried out under the
assumption of uniform rotation.  Differential rotation is directly
observed on the surface on the Sun and measured with
helioseismological analysis in the interior.  A preliminary study of
models with a modest amount of shellular differential rotation
indicates that the details of the graph of $Q'(\hat\omega)$ change
significantly when differential rotation is introduced, while the
typical range of values of $Q'$ obtained is similar.

It is possible that a detailed study taking into account the effects
of stellar evolution (including spin evolution) and differential
rotation may be required to account for the tidal evolution of binary
stars.

\subsubsection{Low-mass binary stars}

Most binary stars do not have equal-mass companions. In the solar
system, the satellites of gas giants generally dominate the
circularization process because they have much lower values of $Q'$
than their host planets and are more intensely perturbed
\citep{GS66,MD99}.  We now consider the contribution of a lower-mass
companion on the circularization process for binary stars.

We have computed the rates of tidal dissipation via inertial and Hough
waves for a $0.5\,M_\odot$ main-sequence star of age $5\,\mathrm{Gyr}$
(Fig.~\ref{f:0.5msun_10d}).  This star is less centrally condensed
than a $1\,M_\odot$ star and the fractional radius occupied by its
radiative zone is smaller.  The contribution of Hough waves is
generally smaller than that of inertial waves unless the tidal forcing
frequencies are outside the range of inertial waves.  Owing to the
different internal structure, the magnitude and frequency-dependence
of $Q'$ in this low-mass stellar model differ from that of a
$1\,M_\odot$ star.  However, when we take into account the mass-radius
relation, the theoretically determined circularization timescale, in
the frequency range of inertial waves, is a generally decreasing
function of the stellar mass.

The results of these models can be directly compared with some
observed systems.  The dataset used to derive a circularization period
for the Hyades includes Johnson 311 \citep{GGGZ85,MM05}.  This system
has a period of 8.5~d and a negligible eccentricity.  The mass of both
of its two components is 0.5 $M_\odot$ which make it an ideal system
for a case study of low-mass stars.  From equation~(\ref{te}), we find
that each star must have $Q'\la1.8\times10^5$ in order for the
circularization timescale to be less than the age of the Hyades
cluster (600~Myr).  This observationally calibrated value of $Q'$ is
more than two orders of magnitude smaller than that computed for the
efficiency of tidal dissipation through both inertial and Hough waves.
The discrepancy here is larger than that found for $1\,M_\odot$ stars.
However, \citet{ZB89} argued that the circularization of these systems
can occur during the Hayashi phase.

\subsubsection{Close-in brown dwarfs and gas giant planets}

All the theoretical models presented here are for uniformly rotating
solar-type stars.  These models are complementary to the models we
presented in Paper~I for gas giant planets.  In our previous analysis,
we showed that the presence of a solid core in gas giants can lead to
a rich spectrum of features in the frequency-dependence of $Q'$. Since
there are extended ranges in the core size and spin periods of gas
giant planets, the dispersion in their values of $Q'$ may be large.
Nevertheless, the typical magnitude of $Q'$ for gas giants is
comparable to that for solar-type stars.

It is useful to make a parallel comparison between the theoretically
determined and observationally inferred values of $Q'$.  Owing to the
technical challenges of the precision measurement of radial
velocities, it is difficult to reduce the uncertainties in planetary
orbital eccentricity to less than about 0.05.  Many planets with
periods less than 5~d have measured eccentricities less than 0.05.  There
are also supposedly eccentric planets within this period range.
However, the apparent eccentricity measurement in such cases may be
due to the perturbation of the host star's motion by additional
planets in the system.

Finally, we note that the brown dwarf companion HD~41004Bb has a
measured eccentricity of 0.08 despite its period of only 1.33~d.  The
mass ratio of the system is about 0.05.  The observed eccentricity can
be preserved over the 1.6~Gyr age of the system provided both objects
have $Q'\ga10^7$.  The value $Q'_*\ga3\times10^7$ required for the
preservation of this brown dwarf implies that the M~dwarf is unlikely
to have been effective in damping the eccentricity of this system (see
the discussions in Section~\ref{s:41004a}).  The lower limit on $Q'_p$
required for the brown dwarf not to damp the eccentricity is an order
of magnitude above that derived for the Jovian planet considered in
Paper~I.  The weak dissipation in this case may be due to the
different internal structure of brown dwarfs.  Hough waves cannot be
excited in these fully convective bodies.  Without the presence of a
core, inertial waves take the form of smooth global modes and are
excited only in the vicinity of particular resonant frequencies
\citep{W05a,W05b}.

\section{Summary and discussion}

In this paper, we have calculated the excitation and dissipation of
low-frequency tidal oscillations in uniformly rotating solar-type
stars and investigated the consequences for the circularization of
binary stars and the orbital migration of close planetary companions.
Our objectives were, first, to try to account for the efficiency of
tidal dissipation inferred from the circularization periods of binary
stars and, second, to try to resolve the paradox that gas giant
planets with very short periods can retained despite their intense
tidal interaction with their host stars.

The detailed analysis presented here is an extension of our previous
investigation.  In Paper~I, we studied the excitation and dissipation
of inertial waves and Hough waves in the convective interior and
radiative surface layers of gas giant planets.  In the present paper
we have examined solar-type stars with convective envelopes and
radiative interiors.  Again, our calculation allows for two avenues of
tidal dissipation.  The disturbance excited in the convective zone,
which takes the form of inertial waves if the tidal frequency is less
than twice the spin frequency, is dissipated by turbulent viscosity.
Hough waves are also excited at the interface between the radiative
and convective regions and propagate towards the center of the star.
We calculate the associated dissipation rate under the assumption that
they do not reflect coherently from the center of the star.

The radiative zone presents a reflecting barrier to inertial waves
because the stable stratification strongly inhibits radial motions of
low frequency.  In this way the radiative interior of the star has a
similar role to the solid core in the planetary model.  When the
inertial waves propagate within a spherical annulus (rather than a
full sphere) the tidal response is spatially intricate and enhanced
dissipation occurs over extended ranges of frequency.  The value of
$Q'$ can be reduced by up to four orders of magnitude when the
Coriolis force is taken into account.

The values of $Q'$ derived for the binary circularization problem in
the case of the present Sun with a spin period of 10~d
(Fig.~\ref{f:1msun_10d}) may still be an order of magnitude too large
to account for the observed circularization periods of solar-type
binary stars.  However, the sensitivity of these results to the tidal
and spin frequencies, the stellar model, and possible differential
rotation means that the mechanisms we describe might be adequate to
explain the observations when a detailed evolutionary study is carried
out.

The dependence of $Q'$ on the tidal and spin frequencies potentially
resolves an outstanding issue with regard to the survival of gas giant
planets with very short-period orbits around solar-type stars.  These
planets cannot supply angular momentum through tidal transfer at a
sufficient rate to compensate for the spin-down of the host stars due
to magnetic braking.  The spin frequencies of these stars are
comparable to that of the Sun, which is at least an order of magnitude
smaller than the tidal forcing frequency.  Consequently inertial waves
cannot be excited.  In addition we estimate that the Hough waves
excited by a hot Jupiter in the present Sun are unlikely to undergo
nonlinear damping at the center of the star.  If the Hough waves
reflect coherently from the center and form global modes with narrow
resonances, their contribution to orbital migration may be negligible.
These very short-period planets can therefore be preserved despite
their proximity to their host stars.  Based on our models, we infer
that gas giants with periods less than 1~d period may be preserved
around solar-type stars.  However, as conditions at the center of the
star evolve, nonlinearity may set in at a critical age, resulting in a
relatively rapid inspiral of the planet.

It would be valuable to conduct numerical simulations of gravity (or
Hough) waves approaching the center of a star to determine at what
amplitude the reflection of the wave is affected by nonlinearity
(cf.~Appendix~\ref{s:center}).  Unfortunately, currently available
stellar models do not agree on the value of $dN/dr$ at the center,
which is of critical importance.

Recently, a number of new planetary transit candidates were announced
with periods less than 1~d in some cases \citep{S06}.  If confirmed,
which seems difficult, these systems would present interesting
constraints on theories of tidal evolution.

\acknowledgments

We are very grateful to Shulin Li and Christopher Tout for providing
stellar interior models and assisting with their interpretation.  This
work was supported by NASA (NAGS5-11779, NNG06-GF45G, NNG04G-191G,
NNG05-G142G), JPL (1270927), NSF (AST-0507424), IGPP, and the
California Space Institute.

\appendix

\section{Gravity waves approaching the center of the star}
\label{s:center}

Low-frequency gravity waves of small amplitude in a non-rotating star
are described by the differential equation \citep[e.g.][]{GD98}
\begin{equation}
  \f{d}{dr}\left[\f{1}{\rho}\f{d}{dr}(\rho r^2\tilde\xi_r)\right]=
  \ell(\ell+1)\left(1-\f{N^2}{\omega^2}\right)\tilde\xi_r,
\end{equation}
where
$\xi_r=\mathrm{Re}[\tilde\xi_r(r)Y_\ell^m(\theta,\phi)\exp(-i\omega
t)]$ is the radial displacement, $\rho(r)$ is the density and $N(r)$
is the \bvf.  Very near the center of the star, the density and
pressure are nearly uniform while $g$ is linear in $r$.  In this
region $\rho$ can be regarded as constant while $N=Cr$ is linear in
$r$.  Let $x=N/\omega=(C/\omega)r$.  Then the regular solution of the
above equation for $\ell\ge1$ is
\begin{equation}
  \tilde\xi_r\propto x^{-3/2}J_\nu(\sqrt{\ell(\ell+1)}x),
\end{equation}
in terms of the Bessel function of order
$\nu=\ell+{\textstyle\f{1}{2}}$.  (The second solution, involving
$Y_\nu$, diverges as $r\to0$.)  In particular, for $\ell=2$,
\begin{equation}
  \xi_r=\f{5A\omega}{8C}x^{-4}\left[6^{-1/2}(1-2x^2)\sin(6^{1/2}x)-
  x\cos(6^{1/2}x)\right]\sin^2\theta\cos(2\phi-\omega t),
\end{equation}
where $A$ is a dimensionless amplitude.  Since the unperturbed entropy
profile is proportional to $r^2$, the perturbed profile is
proportional to $r^2-2r\xi_r$.  The wave overturns the stratification,
according to linear theory, if the quantity
\begin{equation}
  \f{1}{r}\f{\p}{\p r}(r\xi_r)>1.
\end{equation}
The maximum value of this quantity occurs at $r=0$, $\theta=\pi/2$,
and $2\phi-\omega t=0$, and is equal to $A$.  Therefore $A$ is a
measure of the maximum nonlinearity of the wave.  If $A<1$ the entropy
profile is never inverted.  [In this case the wave may still be
disrupted by parametric instabilities, as noted by \citet{GD98}.]

The energy flux in the incoming part of the wave is
\begin{equation}
  F=\f{5\pi A^2}{72\sqrt{6}}\f{\rho\omega^8}{C^5}.
\end{equation}
Equating this to the energy flux in gravity waves excited at the
interface between the radiative and convective zones in a non-rotating
solar model, we estimate
\begin{equation}
  A\approx2500e\left(\f{P}{\mathrm{d}}\right)^{1/6}
\end{equation}
for the largest eccentricity tide involved in the circularization of
a binary with two synchronized $1\,M_\odot$ stars, while
\begin{equation}
  A\approx320\f{M_p}{M_*}\left(\f{P}{\mathrm{d}}\right)^{1/6}
\end{equation}
for the tide excited in a slowly rotating $1\,M_\odot$ star by a
planet of mass $M_p$ and orbital period $P$.  The first estimate is
similar to that made by \citet{GD98}, and implies that the waves
always achieve nonlinearity in the binary circularization problem if
the eccentricity is measurable.  (This remains true when the
Coriolis force is taken into account.)  The second estimate implies
that the waves excited by hot Jupiters are of significant amplitude
but probably do not achieve sufficient nonlinearity to disrupt the
reflection.

These results depend strongly on the stellar model.  For stars of mass
$0.5~M_\odot$ and age $5~\mathrm{Gyr}$ we find
\begin{equation}
  A\approx2.7e\left(\f{P}{\mathrm{d}}\right)^{1/6}
\end{equation}
for binary circularization and
\begin{equation}
  A\approx0.34\f{M_p}{M_*}\left(\f{P}{\mathrm{d}}\right)^{1/6}
\end{equation}
for hot Jupiters.  Therefore even the binary circularization problem
is unlikely to give rise to nonlinear tides if both components have
$0.5~M_\odot$.  The main effect causing this difference is that
$A\propto(dN/dr)_c^{5/2}$ and $(dN/dr)_c$ is about 7~times larger in
the present Sun than in the $0.5~M_\odot$ star.  In general
$(dN/dr)_c$ increases with stellar mass and age, but its value is not
very accurately determined by current models of stellar evolution.

{}

\end{document}